# Photoprotected spin Hall effect on graphene with substrate induced Rashba spin-orbit coupling

Alexander López
*Escuela Superior Politécnica del Litoral, ESPOL, Departamento de Física,
Campus Gustavo Galindo Km 30.5 Vía Perimetral, P. O. Box, 09-01-5863, Guayaquil, Ecuador*

Rafael A. Molina
*Instituto de Estructura de la Materia, CSIC, Serrano 123, 28006 Madrid, Spain*

We propose an experimental realization of the Spin Hall effect in graphene by illuminating a graphene sheet on top of a substrate with circularly polarized monochromatic light. The substrate induces a controllable Rashba type spin-orbit coupling which breaks the spin-degeneracy of the Dirac cones but it is gapless. The circularly polarized light induces a gap in the spectrum and turns graphene into a Floquet topological insulator with spin dependent edge states. By analyzing the high and intermediate frequency regimes, we find that in both parameter limits, the spin-Chern number can be tuned by the effective coupling strength of the charge particles to the radiation field and determine the condition for the photoinduced topological phase transition.

## I. INTRODUCTION

The first proposal for a system displaying the Quantum Spin Hall Effect was made by Kane and Mele by including the effect of spin-orbit interactions on graphene monolayer[1,2]. In the Quantum Spin Hall state there is a vanishing charge Hall conductance but a quantized spin Hall conductance. This proposal gave rise to the whole field of topological insulators, a new state of matter that presents a gap between the valence and conduction bands in the bulk but that supports the presence of gapless edge states at the sample boundaries[3,4]. Topological insulators and Quantum Spin Hall systems present interesting potential applications to spintronics and quantum computation due precisely to the topologically protected edge states[5]. In the Quantum Spin Hall Effect the edge states are protected by charge conservation symmetry while for topological insulators the time-reversal and charge conservation symmetries are needed.

It was soon realized that pristine graphene was not a good candidate to observe experimentally this new effect due to the smallness of the gap induced by the intrinsic spin-orbit coupling and extrinsic Rashba spin-orbit interaction, which were derived by using first principle calculations and were found to be of the order of $\mu$eV[6,7]. Yet, recent experimental results[8–11] and theoretical predictions[12–15] have found that the effective spin-orbit coupling in monolayer graphene can be greatly enhanced to obtain values as large as 100meV. This is achieved by proximity effects with atoms such as Au, Ni, Pb, Ir or Co. Alternatively, the graphene monolayer can also be laid over a layer of a transition-metal dichalcogenide or by proximity to a superconductor, such as $Bi_2Se_3$, allowing the generation of an intrinsic bandgap above room temperature,[16,17]. The main advantage of this proximity induced increase in the spin-orbit strength is that the high mobility of graphene is preserved. Morever, topological insulators have been proposed and observed in semiconductor HgTe quantum wells[18,19]. The experimental observation of topological insulators started a whole new field in condensed matter physics with the prospect of using these new states of matter in spintronics[20] and quantum computation applications[21]. Since these first observations the field of topological quantum matter has increased exponentially with new additions to the family of materials possessing topological properties like 3D topological insulators[22] or topological semimetals[23,24]. More recently, Ezawa[25] has found an exactly solvable model of an f-wave topological superconductor on the honeycomb lattice by considering the role of the Hubbard interaction. Their exactly solvable condition is found to be the emergence of perfect flat bands at zero energy.

A new avenue of research has been the study of Floquet topological insulators where the topological phases of the system can be induced and controlled by an external time-periodic field[26–37]. Graphene was also the first material in which these new systems where proposed[38] but the experimental confirmations came in photonic waveguides arranged in a honeycomb structure[39] and in the three-dimensional topological insulator $Bi_2Se_3$[40,41]. While some works have focused on the tunability of the photoinduced bandgaps,[42] other proposals for the realization of static and photoinduced topological phases have been put forward in silicene[43–45] where, in the driven scenario a single-Dirac cone phase has been shown to appear[46,47]. One interesting result reported in reference[48] is that by including a classical small oscillating lattice distortion (i.e. a phonon mode), which is described as a small time-dependent deformation of the hopping parameter, a nontrivial boundary charge current would emerge in bulk monolayer graphene. Novel photoinduced effects in monolayer transition metal dichalcogenides, such as $WS_2$ have also been reported in reference[49] where it is shown that the driven topological phase is a direct consequence of their intrinsic three-band nature near the band-edges. In addition, these photoinduced Floquet topological phases have also been recently reported in black phosporus[50] where it is found that it exhibits a photon-dressed Floquet-Dirac semimetal state, which can be continuously tuned by changing the direction, intensity, and frequency of the incident laser. Thus, the use of monochromatic laser irradiation does indeed allow for interesting topological scenarios in two-dimensional systems. For instance, the authors of[51] have shown that due to the competition of staggered lattice potential and photon dressing in epitaxial graphene, a threshold value of light intensity is necessary to realize a Floquet topological insulator.

In this paper, we propose a different mechanism for the exper-imental realization of a controllable Quantum Spin Hall effect in graphene. A perpendicular electric field or a substrate breaking inversion symmetry can induce a Rashba spin-orbit coupling in graphene[52]. This coupling can, in principle, be controlled by the intensity of the electric field or the type of substrate but does not fully open a gap in the band structure of graphene[3]. However, in the spirit of the work by Oka and Aoki and the work on Floquet topological insulators[26,38] we propose to open the gap using circularly polarized monochromatic light in the appropriate frequency range. The gap can also be controlled by the parameters of the field allowing the experimental tuning of the effect. We examine some exact analytically solvable parameter regimes and also give a full numerical account of the photo-induced topological phases. The main result is that at intermediate values of the effective light-matter coupling strength, the system undergoes a transition from a nontrivial to a trivial topological phase which is driven by both the radiation field and the Rashba spin-orbit interaction strength. Our results show that the introduction of the spin non-conserving interactions can be exploited to obtain another control parameter to drive the photoinduced topological phases. The structure of the remainder of the paper is as follows. In the next section we present the model and describe the main results based on different analytic approaches. Then, we present a discussion of these results. Finally, we give some concluding remarks and outlook.

## II. MODEL

We consider the changes induced by circularly polarized light in the band structure of monolayer graphene with a perpendicular electric field or lying over a substrate. The substrate induces a homogeneous Rashba spin orbit-interaction on the charge carriers over graphene samples. Measurements of a Rashba coupling constant of 225 meV have been reported on graphene over a substrate of Ni(111)[8,53].

We describe the monochromatic light through a time-periodic dependent vector potential $\vec{A}(t) = \mathcal{E}/\omega(\cos\omega t, \sin\omega t)$ with $\omega$ its frequency, and $\mathcal{E}$ the amplitude of the electric field. In momentum space, and near the **K** point, the time-dependent Hamiltonian reads,

$$H(\vec{p}, t) = v_F \mathbb{I}_s \otimes \vec{\sigma} \cdot [\vec{p} + e\vec{A}(t)] + \lambda(s_x \sigma_y - \tau_z s_y \sigma_x), \quad (1)$$

where $v_F \sim 10^6 m/s$ is the Fermi velocity of the charge carriers in graphene and $\lambda$ the Rashba coupling constant that will depend on the substrate or can be controlled experimentally through a perpendicular static electric field. In addition, $s_j$ ($j = x, y$) describe the Pauli matrices and $\mathbb{I}_s$ is the $2 \times 2$ identity matrix in the real spin degree of freedom, whereas $\vec{\sigma} = (\tau_z \sigma_x, \sigma_y)$ is a vector of Pauli matrices in the pseudospin degree of freedom, where $\tau_z = \pm 1$ refers to the K ($\tau_z = 1$) and K' ($\tau_z = -1$) valley degree of freedom, respectively. In addition, $\vec{p} = (p_x, p_y)$ is the momentum measured from the $\vec{K}$ point. Please, note that this Hamiltonian is a $4 \times 4$ k-dependent matrix in the space of the sub-lattices $A$ and $B$ of the bipartite honeycomb structure of graphene with the spin up and down of the electron. From now, unless explicitly stated we focus on the $\tau_z = +1$ sector.

In matrix form, at the K point, the Hamiltonian in Eq. (1) has a time-independent contribution

$$H_0(\vec{p}) = \begin{pmatrix} 0 & v_F p e^{-i\phi} & 0 & 0 \\ v_F p e^{i\phi} & 0 & 2i\lambda & 0 \\ 0 & -2i\lambda & 0 & v_F p e^{-i\phi} \\ 0 & 0 & v_F p e^{i\phi} & 0 \end{pmatrix}, \quad (2)$$

with $\tan\phi = p_y/p_x$ and a time-dependent contribution

$$V(t) = \begin{pmatrix} 0 & g e^{-i\omega t} & 0 & 0 \\ g e^{i\omega t} & 0 & 0 & 0 \\ 0 & 0 & 0 & g e^{-i\omega t} \\ 0 & 0 & g e^{i\omega t} & 0 \end{pmatrix}, \quad (3)$$

where we have introduced the effective coupling constant $g = ev_F \mathcal{E}/\omega$. In the following we consider a system of natural units where the Fermi velocity of graphene $v_F = 1$ the electron charge $e = 1$ and $\hbar = 1$. The results for the energy bands in the absence of ac-field $\mathcal{E} = 0$ for $\lambda = 0.2$ are shown in FIG.1. This value of $\lambda$ in our natural units corresponds to $\lambda = 119$ meV, well within the achievable experimental values for the spin-orbit coupling[8-11]. Although the Dirac cones are deformed, the system remains gapless.

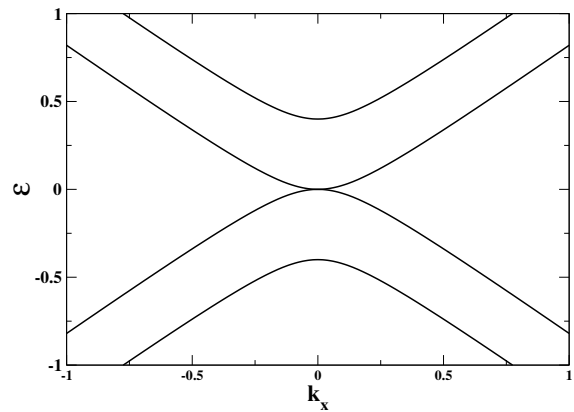

FIG. 1. Band structure of graphene near the **K** point with substrate induced spin-orbit coupling with $\lambda = 0.2\Omega$, Eq. (1).

### A. Gap calculation and analytic solution for $\vec{p} = 0$

First we notice that an exact analytic solution is found at $\vec{p} = 0$, where the Hamiltonian takes the form

$$H(t) = \begin{pmatrix} 0 & g e^{-i\omega t} & 0 & 0 \\ g e^{i\omega t} & 0 & 2i\lambda & 0 \\ 0 & -2i\lambda & 0 & g e^{-i\omega t} \\ 0 & 0 & g e^{i\omega t} & 0 \end{pmatrix}. \quad (4)$$

To diagonalize the Hamiltonian (4) we perform the time-dependent unitary transformation

$$P(t) = \begin{pmatrix} e^{-i\omega t} & 0 & 0 & 0 \\ 0 & 1 & 0 & 0 \\ 0 & 0 & 1 & 0 \\ 0 & 0 & 0 & e^{i\omega t} \end{pmatrix}, \quad (5)$$

that allows us to obtain the quasi-energy spectrum in the first Brillouin zone

$$\varepsilon_{s\eta} = s \frac{\sqrt{2g^2 + 4\lambda^2 + \omega^2 + \eta\sqrt{(\omega^2 - 4\lambda^2)^2 + 4g^2(4\lambda^2 + \omega^2)}}}{\sqrt{2}}, \quad (6)$$

with $s, \eta = \pm 1$ describing the spin and pseudospin degrees of freedom, respectively. For $\lambda = 0$, the spin degeneracy is recovered and the quasi-energies reduce to the well known expressions[38]

$$\varepsilon_{s\eta} = \frac{s}{2}\left(\omega + \eta\sqrt{4g^2 + \omega^2}\right), \quad (7)$$

with $s = \pm 1$ describing the spin degree of freedom, which in this limit is a conserved quantity.

### B. High frequency regime

Although it is possible to find an analytic solution for the Floquet-Schrödinger equation of our problem with momentum $\vec{p} = 0$, there is, in general, no analytic solution for finite momentum. Thus, we consider in the following some approximations where the main dynamical features can be given a semi-analytic treatment. For instance, in the high-frequency regime[46,47] we can apply a sudden approximation useful to describe non-adiabatic processes. Within this frequency regime we approximate the Floquet Hamiltonian as

$$H_F \approx H_0 + \frac{[H_{-1}, H_1]}{\omega} \quad (8)$$

with

$$H_{\pm 1} = \frac{1}{T}\int_0^T dt H(t) e^{\pm i\omega t} \quad (9)$$

where we can approximate the Floquet Hamiltonian as

$$H_F \approx H_0 + \frac{[H_{-1}, H_1]}{\omega} \quad (10)$$

with

$$H_{\pm 1} = \frac{1}{T}\int_0^T dt H(t) e^{\pm i\omega t} \quad (11)$$

Thus, the physically relevant processes are those corresponding to virtual emission and absorption of one photon that in the literature describe so called dressed states. Within this regime one finds explicitly

$$H_1 = g\begin{pmatrix} 0 & 1 & 0 & 0 \\ 0 & 0 & 0 & 0 \\ 0 & 0 & 0 & 1 \\ 0 & 0 & 0 & 0 \end{pmatrix}, \quad (12)$$

and $H_{-1} = H_1^\dagger$. The approximate Floquet Hamiltonian is then

$$H_F = \begin{pmatrix} -\Delta & v_F p e^{-i\phi} & 0 & 0 \\ v_F p e^{i\phi} & \Delta & 2i\lambda & 0 \\ 0 & -2i\lambda & -\Delta & v_F p e^{-i\phi} \\ 0 & 0 & v_F p e^{i\phi} & \Delta \end{pmatrix}, \quad (13)$$

with $\Delta = g^2/\omega$. Therefore, within this perturbative regime we find the quasi-energy spectrum $\varepsilon_{s\eta} = s\varepsilon_\eta$, with $s, \eta = \pm 1$ representing the spin degree of freedom and

$$\varepsilon_\eta = \sqrt{\left(\sqrt{v_F^2 p^2 + \lambda^2} + \eta\lambda\right)^2 + \Delta^2}. \quad (14)$$

Thus, the driving field introduces a gap in the energy spectrum as it is described in reference[38] where circular polarized radiation field opens exactly the same gap in the energy spectrum. In order to describe the topological properties we find an equivalent $2 \times 2$ reduced Hamiltonian, as follows. The Floquet Hamiltonian is written as

$$H_F = \begin{pmatrix} H & 2i\lambda\sigma_- \\ -2i\lambda\sigma_+ & H \end{pmatrix}, \quad (15)$$

with the $2 \times 2$ Hamiltonian

$$H = \begin{pmatrix} -\Delta & v_F p e^{-i\phi} \\ v_F p e^{i\phi} & \Delta \end{pmatrix}, \quad (16)$$

and

$$\sigma_\pm = \frac{\sigma_x \pm i\sigma_y}{2}. \quad (17)$$

Writing the Schrödinger equation for $H_F$

$$H_F|\Psi\rangle = \varepsilon|\Psi\rangle \quad (18)$$

and putting its eigenstates as

$$|\Psi\rangle = \begin{pmatrix} |u\rangle \\ |d\rangle \end{pmatrix}, \qquad (19)$$

with $|u\rangle$ ($|d\rangle$) the upper (lower) spinor component we obtain, after elimination of the upper component, an effective Hamiltonian for the lower spinor, which apart from an energy shift term, is reading as

$$H_{s\eta} = \Delta_{s\eta}\sigma_z + v_F p(e^{-i\phi}\sigma_+ + e^{i\phi}\sigma_-) \qquad (20)$$

where the effective momentum-dependent gap is given by

$$\Delta_{s\eta} = \frac{\eta\lambda(s\varepsilon_\eta + \Delta)}{\eta\lambda + \sqrt{v_F^2 p^2 + \lambda^2}} - \Delta. \qquad (21)$$

We check that for $\lambda = 0$ we recover the photo-induced gap corresponding to spin degenerate case. For the parameter regime of interest, i.e., $\lambda, \Delta \ll \omega$, we can further simplify the expression for the effective bandgap and approximate it as follows:

$$\Delta_{s\eta} \approx \frac{s\eta\lambda\varepsilon_\eta}{\eta\lambda + \sqrt{v_F^2 p^2 + \lambda^2}} - \Delta, \qquad (22)$$

$$\approx \frac{s\lambda\eta\sqrt{\left(\eta\lambda + \sqrt{v^2 p^2 + \lambda^2}\right)^2}}{\eta\lambda + \sqrt{v_F^2 p^2 + \lambda^2}} - \Delta, \qquad (23)$$

$$\Delta_{s\eta} \approx s\eta\lambda - \Delta. \qquad (24)$$

Thus, as in the Kane and Mele model, the topological invariant is given by

$$C_\beta = \mathbf{sign}(\Delta_\beta). \qquad (25)$$

with $\beta = s\eta$. In this manner, the associated invariant can be controlled both by the Rashba interaction term and effective light-matter bandgap. We remark that although within this regime these two parameters are small compared to the frequencies of interest, they could still have experimentally achievable values that could afford an experimental probe for the realization of the photo-induced topological nontrivial phases described. The phase diagram corresponding to the Chern invariant for the parameter $\beta = s\eta = +1$, i.e. $C_{+1} = C$, is shown in FIG.2, whereas it is clearly seen from equation 25 that the corresponding high frequency Chern invariant $C_{-1}$ satisfies $C_{-1}(\lambda, \Delta) = -C_{+1}(\lambda, -\Delta)$.

### C. Numerical results

In general, the treatment of the dynamical effects would require a numerical treatment which we present in this section. The problem of graphene in the presence of circularly polarized light was tackled by Oka and Aoki in the context of the appearance of a photovoltaic Hall effect[38]. They used the Floquet method[54–56] to solve the Schrödinger equation for the time-periodic Hamiltonian of Eq. (1) with $\lambda = 0$.

$$i\partial_t|\Phi(\vec{k}, t)\rangle = H(\vec{k}, t)|\Phi(\vec{k}, t)\rangle. \qquad (26)$$

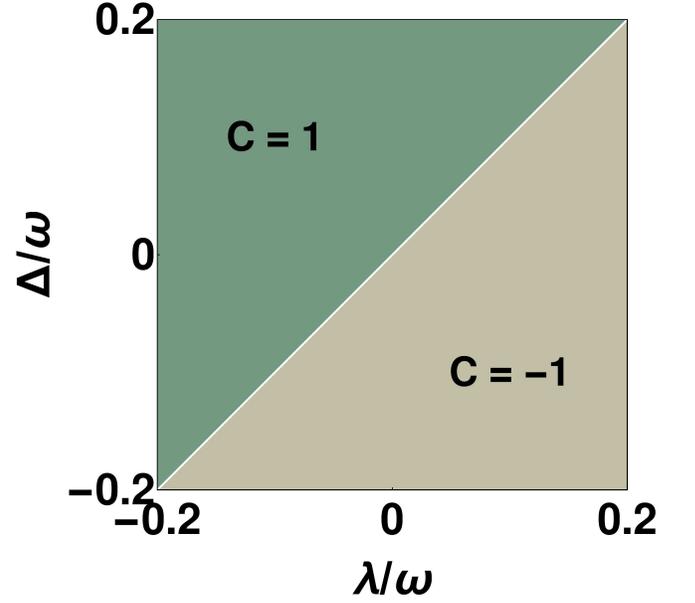

FIG. 2. Phase diagram in the $(\lambda, \Delta)$-parameter space for the Topological invariant $C_\beta$, with $\beta = +1$, which is obtained from the high frequency regime as described in equation (25).

For this purpose one can define an auxiliary hermitian Hamiltonian $\mathcal{H}(t) = H(t) - i\partial_t$, along with the so called Floquet states $|\Psi_\alpha(t)\rangle = \exp(i\varepsilon_\alpha t)|\Phi(t)\rangle$, such that $\mathcal{H}(t)|\Psi_\alpha(t)\rangle = \varepsilon_\alpha|\Psi_\alpha(t)\rangle$, which are periodic functions of time, $|\Psi_\alpha(t + T)\rangle = |\Psi_\alpha(t)\rangle$, and $\varepsilon_\alpha$ are called the quasi-energies, and are the analogous of the quasi-momenta for Bloch electrons in a spatially periodic structure. Since the states $|\Psi_{\alpha n}(t)\rangle = \exp(in\omega t)|\Psi_\alpha(t)\rangle$ are also eigenstates of $\mathcal{H}(t)$ with eigenvalues $\varepsilon_\alpha \to \varepsilon_\alpha + n\omega$, we can work in the *first Brillouin zone* $-\omega/2 \leq \varepsilon_\alpha \leq \omega/2$, with $\omega = 2\pi/T$.

Using the periodic temporal basis $\xi_n(t) = \exp(in\omega t)$, which satisfies

$$\frac{1}{T}\int_0^T \xi_n^*(t)\xi_m(t)dt = \delta_{nm},$$

writing the Fourier expansion

$$\Phi_{\alpha n}(t) = \exp(-i\varepsilon_\alpha t) \sum_{n=-\infty}^{n=\infty} |C_\alpha^{(n)}(\vec{k})\rangle\xi_n(t), \qquad (27)$$

and using the expansion $|C_\alpha^{(n)}(\vec{k})\rangle = \sum_\beta \Phi_{\alpha\beta}^{(n)}|\phi_\beta(\vec{k})\rangle$, Eq.(26) becomes

$$H(\vec{k}, t)\sum_{n=-\infty}^{\infty}\sum_\beta \Phi_{\alpha\beta}^{(n)}|\phi_\beta^{(0)}(\vec{k})\rangle\xi_n^*(t) + \qquad (28)$$

$$\sum_{n=-\infty}^{\infty}\sum_\beta \Phi_{\alpha\beta}^{(n)}|\phi_\beta^{(0)}(\vec{k})\rangle\xi_n^*(t)(\varepsilon_\alpha - n\omega) = 0.$$

Multiplication by $\langle\phi_\gamma(\vec{k})|\xi_m^*(t)$, then integration in the momen-

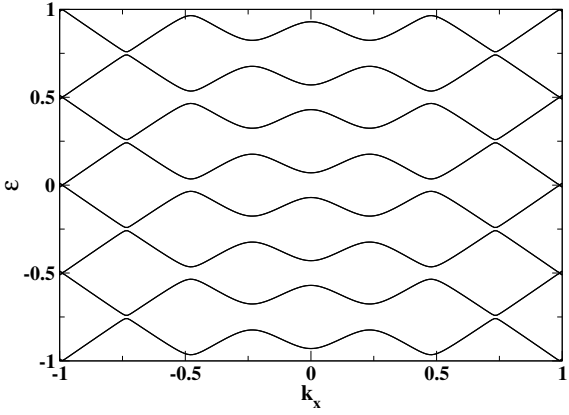

FIG. 3. Band structure of graphene with no substrate induced spin-orbit coupling with $\lambda = 0$ but external field $g = 0.2\omega$ and $\omega = 0.5$.

tum domain, and average over one temporal period, leads to

$$\sum_{n=-\infty}^{n=\infty} \sum_{\beta} [\langle \alpha | H^{(m-n)} | \beta \rangle - (\varepsilon_\alpha - m\omega)\delta_{nm}\delta_{\alpha\beta}]\Phi^{(n)}_{\alpha\beta} = 0. \quad (29)$$

and we have used the simplifying notation $|\alpha\rangle \equiv |\phi_\alpha(\vec{k})\rangle$, and $H^{(m-n)} = 1/T \int_0^T \xi_m^*(t) H(\vec{k},t)\xi_n(t)$.

Then, the quasi-energies $\varepsilon_\alpha$ are eigenvalues of the secular equation.

$$det|H_F - \varepsilon_\alpha| = 0 \quad (30)$$

where $\langle \alpha m | H_F | n\beta \rangle = H^{(m-n)}_{\alpha\beta} + m\omega\delta_{nm}\delta_{\alpha\beta}$.

In the $\lambda = 0$ case the $4 \times 4$ matrices split in $2 \times 2$ blocks and the spin up and spin down sectors can be solved independently. For example, in the case of $\tau_z = 1$ the Floquet Hamiltonian have diagonal blocks

$$H^{mm} = \begin{pmatrix} 0 & k_x - ik_y \\ k_x + ik_y & 0 \end{pmatrix}, \quad (31)$$

while the nondiagonal blocks are

$$H^{mm+1} = \begin{pmatrix} 0 & g \\ 0 & 0 \end{pmatrix}, \quad H^{mm-1} = \begin{pmatrix} 0 & 0 \\ g & 0 \end{pmatrix}. \quad (32)$$

For numerically solving these equations we need to truncate at a certain value of $|m|$.

The band structure of the first Brillouin zone of the quasi-energies is then shown for $g = 0.2\omega$ considering $\lambda = 0$ in Fig. 3. The band structure is periodically repeated as we mount up in energy.

The ac field opens gaps at $\epsilon = \pm\omega/2$ and further gaps separated by the field frequency $\omega$. A gap also opens at $k = 0$ and $\epsilon = 0$. Oka and Aoki showed that this special band structure has the effect of producing a photovoltaic dc Hall current[38].

Similarly we can combine both results to solve the Floquet Hamiltonian in the presence of Rashba spin-orbit coupling.

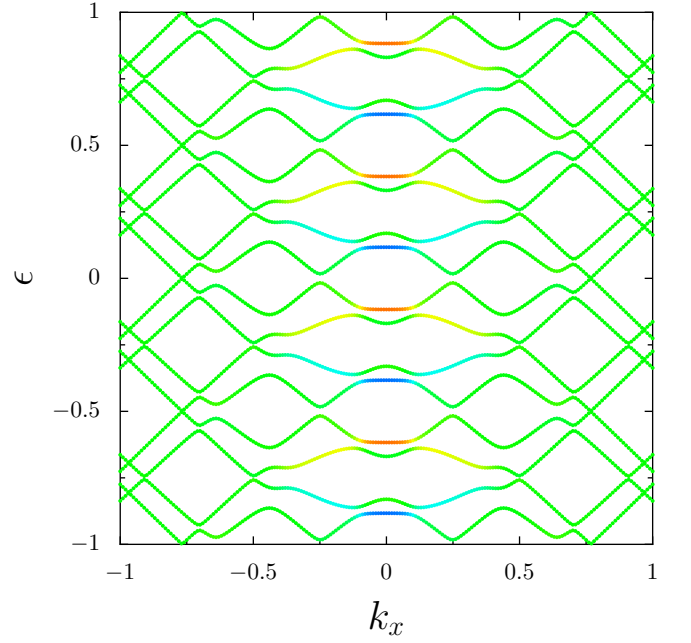

FIG. 4. Band structure of graphene with $\lambda = 0.2\omega$, $g = 0.2\omega$ and $\omega = 0.5$. The color code of the points represent the $S_z$ component of the states. Blue positive, green zero, and red negative.

The Floquet Hamiltonian is now a 4×4 matrix with sectors for up and down spin with the previous structure and connected by terms $H^{mm}_{RO}$ in the diagonal blocks

$$H^{mm}_{RO} = \begin{pmatrix} 0 & 0 & 0 & 0 \\ 0 & 0 & 2i\lambda & 0 \\ 0 & -2i\lambda & 0 & 0 \\ 0 & 0 & 0 & 0 \end{pmatrix}. \quad (33)$$

For comparison to figure **??** we show in **??** the quasienergy spectrum for $\omega = 0.5$, $g = 0.2\omega$ for finite $\lambda = 0.2\omega$ spin-orbit interaction. We notice that spin splitting and additional photoinduced bandgaps appear due to the finite value of *lambda*. The spin texture is shown with a color code indicating the $S_z$ component of the spin. In order to characterize the topological properties of the model with the different parameters we calculate the Chern number $C$ corresponding to the Floquet bands using the Thouless-Kohmoto-Nightingale-den Nijs (TKNN) formula[57].

$$C = \frac{1}{2\pi} \sum_n \int d^2k (\Omega_n)_z. \quad (34)$$

where $n$ labels the occupied bands, and $\Omega_n$ is the Berry curvature

$$\Omega_n = i \left\langle \frac{\partial u_n}{\partial \vec{k}} \middle| \times \middle| \frac{\partial u_n}{\partial \vec{k}} \right\rangle, \quad (35)$$

with $u_n$ denoting the Floquet-Bloch states for band $n$. The filling of the bands in the Floquet case outside of equilibrium depends on the initial state and the way we switch on the external field. However, we will assume that the field is switched on adiabatically so the Fermi energy does not change from the static case.

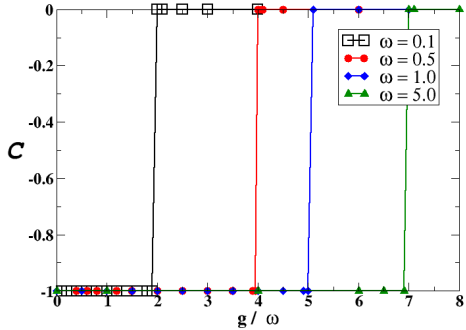

FIG. 5. Numerically calculated Chern number for the case $\lambda = 0.2\omega$ and four values of the frequency $\omega = 0.1, 0.5, 1.0, 5.0$.

The integration in $d^2k$ extends to the first Brillouin zone. In the adiabatic case the Chern number calculation should give an integer value, however in the non-adiabatic case the result can be a real number (in general, smaller than in the adiabatic case). We are calculating the Chern number from the Floquet cyclic states so this is only a geometrical property of the Floquet bands and not a dynamical property and we should always get an integer number. In general, we obtain a transition from Chern number $C = -1$ ($-2$ if we include the symmetric $K$ and $K'$ points) to Chern number $C = 0$ as we increase the amplitude of the external field. This is in contrast to the results of the previous section but the transition value of the coupling constant $g$ is outside the high frequency, low amplitude approximation used in Eqs. (24) and (25). We can see in FIG.5 that the transition is displaced to higher values of $g$ for higher frequencies.

In FIG.6 we can see how the transition is reflected in the band structure. The gap increases with $g$ but for large values of $g$ there appears a band crossing at a finite value of $k$ switching the spin components between occupied and unoccupied bands. At this point, the spin texture is lost. This inversion of the bands marks the transition from the topological insulator phase to a topologically trivial phase.

## III. CONCLUSIONS AND PERSPECTIVES

We have proposed an experimentally feasible model for a Floquet topological insulator realization by means of the interplay among Rashba spin-orbit interaction and monochromatic circularly polarized radiation. We have first shown, via an analytically solvable approach, that the high frequency regime allows for the realization of a nontrivial quantized Chern number whose value depends on both the spin-orbit as well as the effective light-matter interaction strengths. Interestingly, the realization of such nontrivial topological phase has been vali-dated via a full numerical calculation for the intermediate frequency regime, where the radiation field's frequency is either smaller or comparable to the tight-binding hopping parameter which has been used as the energy scale of the problem. Naturally, these results can be extended to other two dimensional materials, such as the transition metal dichalcogenides, where a larger spin-orbit interaction is achievable.

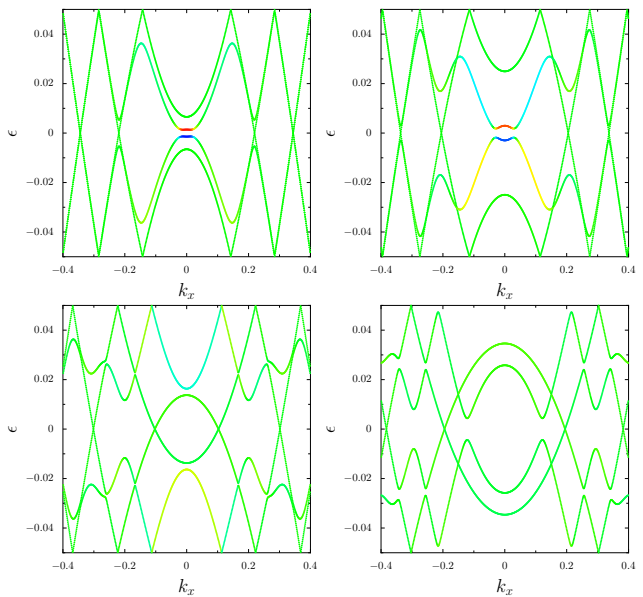

FIG. 6. Band structure of graphene with $\lambda = 0.2\omega$, $\omega = 0.1$, and different values of $g = 0.5\omega, \omega, 2\omega, 3\omega$ from left to right and top to bottom. The color code of the points represent the $S_z$ component of the states. Blue positive, green zero, and red negative. The spectrum changes shape when $g \simeq 2\omega$ as the Chern number goes form $-1$ to $0$ and the gap closes.

Moreover, the role of periodic driving in modulating the geometric nature of quantum states has also been recently put forward for revealing an important physical quantity of interest as the metric tensor, which in turn, can be used to quantify the "distance" among quantum states[58]. In this work it is shown that the diagonal (off-diagonal) elements of the metric tensor are related to the integrated transition rate (excitation rates) in driven systems. This metric tensor approach has also been recently theoretically explored in analyzing the static properties of materials such as borophene[59] and the organic compound $\alpha$-BEDT[61] for which the low energy physics corresponds to so called "tilted" Dirac or Weyl materials. The tilted nature of the energy spectrum can lead to anisotropic energy spectra which, from the interpretation via the metric tensor can be used as a platform for describing analogs of gravitational waves produced by electric fields instead of requiring mass sources and the authors argue that these solid-state "gravitational waves" could be tested by spectroscopic experimental techniques. In addition, the tilted description in Weyl semimetals coupled to electromagnetic fields can be encoded in the metric tensor approach which leads to peculiar behavior of the surface plasmon polariton physics[60]. Among the unusual properties of the surface plasmon polariton with much higher frequency than the bulk plasmon frequency. Although the lattice structure of graphene does not match the required symmetries for the realization of these phenomena within the static or driven scenario it would be interesting to explore the dynamical aspects of the tilted Dirac and Weyl materials subject to light-matter interactions and this could be the subject of future research.

The experimental means that can be suitable for the detection of our theoretical findings should be a combination of two techiques. On the one hand, there is the time-resolved ARPES techique, as has been already implemented in other systems such as the photonic waveguides arranged in a honeycomb structure[39] and in the three-dimensional topological insulator $Bi_2Se_3$ studied in references[40,41]. We would also expect that some recent experiments in the far infrared frequency domain[62] for which $\omega \approx 10 meV$ and values of the electromagnetic field intensities $\mathcal{E} \sim 0.15 MV/m$ could be suitable for validating the reported results. In addition, for the experimental feasibility of addressing an enhanced effective spin-orbit interaction via proximity effects, the time resolved ARPES techique could be suitable accompanied by a setup for the detection of the photoinduced phase transitions by employing the spin-resolved ARPES (SARPES) techique[63–65]. We expect then, that our reported results could motivate the experimental pursue of the photonic manipulation of the spin degree of freedom in graphene-like systems.

*Acknowledgments* This work has been supported by CEDIA via the project CEPRA-XII-2018-06 Espectroscopía Mecánica: Transporte interacción materia radiación. We also acknowledge financial support through Spanish grants PGC2018-094180-B-I00 (MCIU/AEI/FEDER, EU) and FIS2015-63770-P(MINECO/FEDER, EU), CAM/FEDER Project No.S2018/TCS-4342(QUITEMAD-CM) and CSIC Research Platform PTI-001.